\documentclass[twocolumn,secnumarabic, amsmath,amssymb, nobibnotes, aps, prd,floatfix,showkeys, showpacs]{revtex4-1}
\usepackage{siunitx}
\usepackage{graphicx}
\usepackage{hyperref}
\usepackage{color}

\newcommand{\avN}[0]{\langle N \rangle}
\newcommand{\lowt}[1]{ _{\mathrm{#1}}}

\newcommand{\figref}[1]{Fig.#1}

\begin{document}

\title{Area confined position control of molecular aggregates}

\author{Hong Wang$^1$}
\author{ Oleg Buller$^2$}
\author{ Wenchong Wang$^{1}$}
\email[ To whom correspondence should be addressed: ]{wangw@uni-muenster.de, chi@uni-muenster.de}
\author{ Andreas Heuer$^2$}
\author{ Deqing Zhang$^3$}
\author{ Harald Fuchs$^1$}
\author{Lifeng Chi$^{1,4}$}%
\email[ To whom correspondence should be addressed: ]{wangw@uni-muenster.de, chi@uni-muenster.de}
\affiliation{1 Physikalisches Institut and Center for Nanotechnology (CeNTech), Universit{\"a}t M{\"u}nster, 48149 M{\"u}nster, Germany}
\affiliation{2 Institut f{\"u}r Physikalische Chemie, Universit{\"a}t M{\"u}nster, 48149 M{\"u}nster, Germany}
\affiliation{3 Beijing National Laboratory for Molecular Sciences, Organic Solids Laboratory, Institute of Chemistry, CAS, 100190 Beijing, China}
\affiliation{4 Jiangsu Key Laboratory for Carbon-based Functional Materials \& Devices Collaborative,
 Institute of Functional Nano \& Soft Materials (FUNSOM) and Collaborative Innovation Center of
Suzhou Nano Science and Technology, Soochow University, Suzhou 215123, P. R. China}

\begin{abstract}
We report an experimental approach to control the position of molecular aggregates on
surfaces by vacuum deposition. The control is accomplished by regulating the
molecular density on the surface in a confined area.
The diffusing molecules
are concentrated at the centre of the confined area, producing a stable cluster
when reaching the critical density for nucleation. Mechanistic aspects of that control
are obtained from kinetic Monte Carlo simulations.  The dimensions of the position can further be
controlled by varying the beam flux and the substrate temperature.
\end{abstract}
\maketitle

Physical vapor deposition (PVD) describes a technique to condense materials
onto a surface. Typically the materials are vaporized to generate atomic
and molecular beams, and directed onto a substrate in vacuum \cite{Foster199797}. The method
allows for novel architectures with atomic precision control like artificial
heterostructures \cite{Vomiero20077}. Driven by the intensive applications in organic
electronics, functional small molecules have attracted much attention over the
last three decades \cite{Yamada200826, Uoyama2012}. Owing to the superior device performances over
other techniques like spin-coating, the PVD is widely used for functional small
molecule film preparation in both academic researches and industrial
productions \cite{Prantontep200572, Lucas201261}.

The basic growth process of molecules by PVD involves absorption,
diffusion, desorption and nucleation of molecules on the surface. The
nucleation contains the gathering of molecules at specific sites over a
critical size and evolving to dynamically stable clusters \cite{Pimpinelli20145}. In
analogy to inorganic atoms, the organic molecules are found preferably to nucleate at
defects, step edges, and aggregate together when a sufficient number of molecules is close together
\cite{Maksymovych2005109, Glowatzki200776,Wagner2013110, Pratontep200469}. In order to generate
regular-spaced nanostructures with this general approach, e.g., train-relief patterns \cite{Brune1998} or hydrogen-bonded
surface networks \cite{Madueno2008} can be used. Recent examples to generate position control, color tuning with two dyes
and improved carrier mobility for the organic field effect transistors can be found in
\cite{Wang200798, Wang200921, Wang201022, Wang201410}. Whereas the position control itself is quite insensitive to the chosen
molecules, the specific properties of the individual aggregates naturally reflect the details of the individual molecular
interactions; see, e.g. \cite{Kowarik200696, Bommel2014}.

Whereas the methods to obtain position control, used so far, rely on the presence of surface-supplied nucleation centers,
we present a method which works without specific nucleation centers.
Furthermore, it can generate regular structures on the micrometer scale.
Due to the randomness of the trajectories of the
molecules this may seem to be very difficult \cite{Wang201245}. The key idea is generate an initial inhomogeneous density
profile of the molecules. Via combination with kinetic Monte Carlo simulations we also succeed to obtain a mechanistic
understanding of the experimentally observed effects. Since the proposed concept  does not depend on any molecular details,
the simulations are performed for a minimum representation of the system.

\begin{figure}[b]
  \centering
  \includegraphics[width=0.5\textwidth]{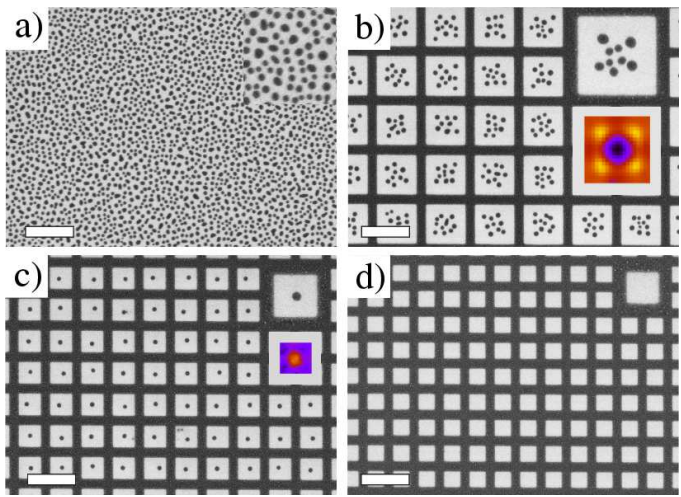}
  \caption{\label{fig:fig1}
    SEM images of NPB deposited on a) bare, b) \SI{4.0}{\micro\meter} c)
    \SI{2.2}{\micro\meter} and d) \SI{1.8}{\micro\meter} Au grid patterned SiO$_2$ surface;
    inset: magniﬁed view and in b) and c) the corresponding color coded island
    size distribution from the simulations (yellow high, black low value). The scale bar here is
   \SI{5}{\micro\meter}.
  }
\end{figure}

Previously, vicinal surfaces were applied to create molecule density distribution, which the step edges act as the sinks
for adatoms \cite{Ranguelov200775, Altman200266} However, the methods are limited to specific substrates such as single crystals and lead to no
ordering of the aggregates owing to random presence of step edges on the surface. In our case, we experimentally patterned
the SiO$_2$ with Au grids by standard electron
beam lithography  \cite{Wang200798}. The Au grid consists of two orthogonal line arrays
with a width of \SI{1}{\micro\meter}. The spacing varies from \numrange{1.6}{5.0}
 \si{\micro\meter}. For the
functional molecule, we choose N, N'-bis(1-naphthyl)- N, N'-diphenyl-1,1'-biphenyl'-4,4'-diamine
(NPB, a molecule widely used for organic light emitting diodes)  \cite{Forsythe199873, Supp_mat1}. Fig.  \ref{fig:fig1} a)
shows scanning electron microscope (SEM) images of NPB deposited on a bare SiO$_2$ surface. In contrast,
in Figs. \ref{fig:fig1} b)-d) an Au grid with a spacing of
 \SI{4.0}{\micro\meter}, \SI{2.2}{\micro\meter}, and \SI{1.8}{\micro\meter} have been employed, respectively.
The molecules were deposited at a substrate temperature of \SI{200}{\celsius} to
ensure the diffusion of molecules over the surface and at a deposition rate of \SI{0.2}{\nano\meter/\minute}
for 20 minutes.
The SEM images were taken in second electron mode with an inlens detector, using \SI{3}{\kilo\volt}
accelerating voltage to avoid damage of the organics.
After the deposition, the sample was cooled down to room temperature and
characterized ex situ by SEM to view the position of the NPB aggregates
(dark points in the images).

As expected and observed extensively, without the Au grid the NPB islands are distributed in
a random fashion \cite{Ala-Nissila200251}. In contrast, for the \SI{4.0}{\micro\meter} Au grid, several
molecular islands are present in the centre of the grid. Notably the number of islands
is quite uniform ranging from 9 to 11 in each cell. The largest islands tend to be closer to the four edges of the Au square.
The number of the islands in the cells decreases with the size. When the grid size decreases to \SI{2.2}{\micro\meter}, only
one island is present in the centre of the cell, leading to the number
and position control of the molecule aggregation. With optimization of
growth conditions, most remarkably, more than 95 \% of all cells contain exactly one island. This corresponds to
a high-quality growth control.
As the grid size further decreases down to \SI{1.8}{\micro\meter}, shown in Fig. \ref{fig:fig1} d), all
molecules can diffuse to the Au, resulting in patterned growth of organic
molecules \cite{Ala-Nissila200251}. The volume of molecules on SiO$_2$ with different
grid sizes was calculated, as shown in Fig. S2. The volume on SiO$_2$ increase with the
grid size, showing the continuous lost of control of the patterns.

\begin{figure}[t]
  \centering
  \includegraphics[width=0.42\textwidth]{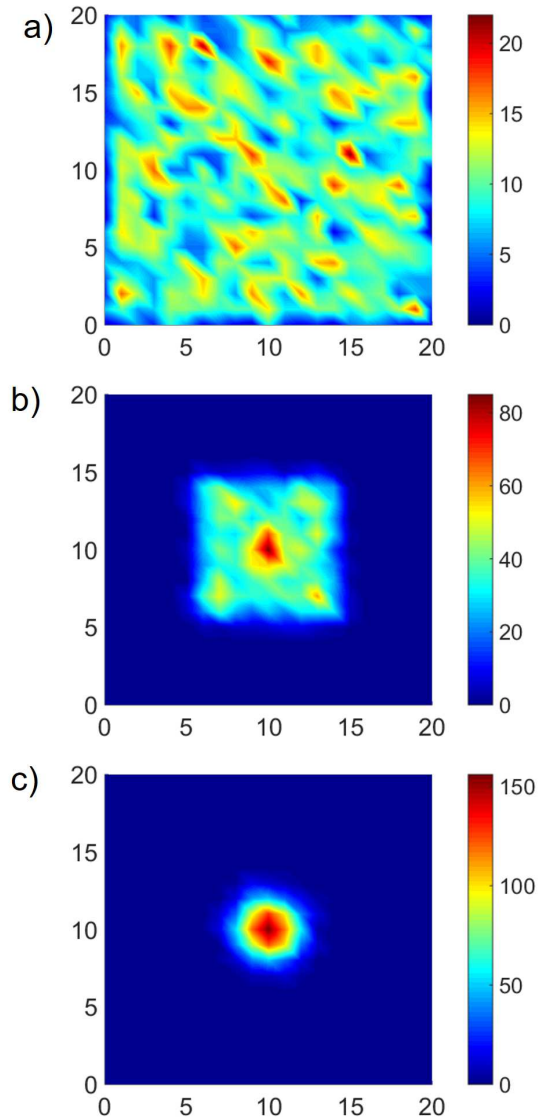}
  \caption{\label{fig:fig2}
    Histogram of NPB islands position distribution \mbox{in a)} \SI{2.2}{\micro\meter} grid
    artificially added on bare SiO$_2$, for comparison, b) \SI{4.0}{\micro\meter} Au grid and
    c) \SI{2.2}{\micro\meter} Au grid.
  }
\end{figure}

Quantitatively, we analyzed the position of each island on the samples
shown in Fig. \ref{fig:fig1} a)-c). The analysis was
performed by dividing each grid
cell into a 20 by 20 mesh which generates an X-Y coordinate system.
For comparison, a virtual grid in size of \SI{2.2}{\micro\meter} is artificially added to
the sample of the unpatterned SiO$_2$. By mapping the centre of mass grid cell
to the mesh, we get the position of each island. In total we
measured and counted 235 virtual grid cells with around 4000 islands for
bare SiO$_2$, 300 with 3100 islands for the grid size of \SI{4.0}{\micro\meter},
and 1800 with 1716 islands
for the grid size of \SI{2.2}{\micro\meter}. The island position distribution is shown in
Fig. \ref{fig:fig2} a)-c).

Naturally, the molecular islands on bare SiO$_2$ surface display a uniform
distribution in the virtual grid, reflecting a random location of the
aggregates (Fig. \ref{fig:fig2} a). For the grid with the size of \SI{4.0}{\micro\meter}, the island
distribution also displays the square symmetry, but is shrunk to a smaller size (Fig. \ref{fig:fig2} b). As the grid size further
decreased to \SI{2.2}{\micro\meter}, the island position distribution shrinks to a point,
located in the centre. Thus, also the position control is excellent (Fig. \ref{fig:fig2} c).

To obtain information about the mechanism of position and number
control, we performed kinetic Monte Carlo simulations with the same surface setup as in the experiment.
We used a three-dimensional discrete model, similar to model described in \cite{lied:2012}. One quadratic simulation box represents one
cell with periodic boundary conditions in $x$- and $y$- direction. We used cell sizes (gap between gold stripes)
of $L=46a$, $L=80a$ and $L=146a$ to analyze the scaling behavior of island formation in the centre of the cell.
The variable $a$ is the distance between two lattice points. The width of the
gold array for all cells is 20$a$. The system contains three particle
types, the substrate particle $s$, the gold particle $g$ and the deposited particle $p$. Each
particle type can occupy one lattice site. The
gold particles are fixed during the simulation and are incorporated in the lowest
substrate plane to avoid step edge barriers. The interaction energies (relative to $k_{\mathrm{B}} T$) $\varepsilon_{sp} = 0.3$,
$\varepsilon_{gp}=1.3$ and
$\varepsilon_{pp}=1.0$ have been chosen to mimic the situation in the experiment. Specifically, it has been shown
that the growth of NPB on single gold stripes can be reproduced very well \cite{lied:2012}.
Nonetheless the absolute values of the interaction parameters are not decisive for the phenomena,
crucial is only the condition $\varepsilon_{sp} < \varepsilon_{gp}$.
The simulation starts with no particles on the substrate. Per
Monte Carlo step, every particle on the substrate makes a Monte Carlo move according to
 the Metropolis criterion \cite{Metropolis53} and finally $n$ particles are added to the system. $n$ is
Poisson distributed with the mean value $\bar{n}$, which is related to the average flux by
 $F = \bar{n}/(A \Delta t)$. Here
$A$ is the surface size and $\Delta t$ the corresponding Monte Carlo time step.
Particles are put directly on the surface, but cannot detach from it during the simulation.
The simulation finishes, as in the experiment, when two monolayers (ML number of particles per full surface coverage) are
deposited on the surface. The presented data were obtained from 2000 independent simulations
with an average flux of
\num{5.8d-6}$/(a^2 \Delta t)$.

We start by reporting the average
projected cluster size distribution for the cell sizes of 80$a$ and 146$a$, where we get on average
one and eight clusters, respectively; compare Fig. \ref{fig:fig1} b) and c).
Thus, these cell sizes thus can be approximately related to the \SI{2.2}{\micro\meter} and \SI{4.0}{\micro\meter}
cells in the experiment. We used the projection of the three-dimensional deposited
particle distribution onto the surface $xy$-plane $P(x_i,y_j)$, with $x_i$ and $y_j$ as the
discrete lattice position in the cell. If the position $(x_i,y_j)$
is occupied by a deposited particle we choose $P(x_i,y_j) = 1$, otherwise $P(x_i,y_j) =0$. The
projected field was used to determine the two-dimensional size and the centre of
mass of each cluster. Only if more than $b=4$ particles in $P(x_i,y_j)$ stick together, they are
considered to be a cluster. The choice of $b$ reflects the critical nucleus size as estimated in analogy
to \cite{Mues}. The result, however, is insensitive to the specific value of $b$, because the typical
size of clusters is by far larger so that the identification of a cluster is insensitive to this choice.
To get the averaged distribution, we divided the centre of
mass coordinates into a $20a \times 20a $ mesh and calculated the average of the cluster
sizes for each mesh cell. This color coded plot is included to Fig. \ref{fig:fig1} b)
and \ref{fig:fig1} c).
In the case of lattice size of 80$a$ the biggest islands are in centre of the cell, whereas
for the lattice of 146$a$ they are near the corners of the cell.
Both observations fully agree with the experiment.

The key advantage of simulations is to get information about the time evolution.
For this purpose,
we analyzed the particle density distribution $\rho(x_i,y_j)$ of the cells
as given by the ensemble
average of $P(x_i,y_j)$ of the cells as a function of time. To elucidate the molecular density {\it before} the
nucleation event we explicitly identified at each time step those simulation runs without prior cluster formation.
Of course, in this {\it cluster-free subensemble} the number of contributing simulations decreases with time.
Specifically, $\rho^*$ denotes the molecular density in the cluster-free subensemble in the centre of the cell.
To increase the statistics of $\rho^*$ we averaged over same lattice points in the center of the cell.

\begin{figure}[t]
  \centering
  \includegraphics[width=0.39\textwidth]{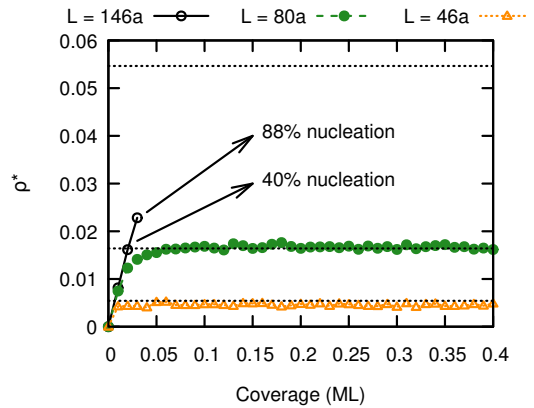}
  \caption{\label{fig:fig3}
     Time dependence of $\rho^*$ for the three different system sizes $L=46a, L=80a$,
     $L=146a$ and the corresponding analytical stationary value of $\rho^*\lowt{stat}$ (broken lines).}
\end{figure}

\begin{figure}[t]
  \centering
  \includegraphics[width=0.39\textwidth]{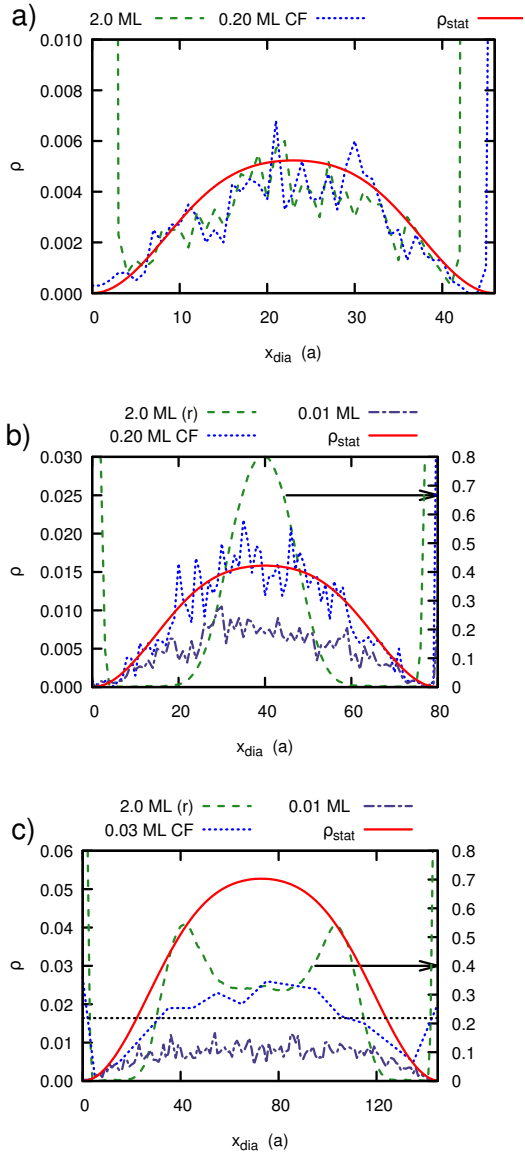}
  \caption{\label{fig:fig3b}
   (a) Time evolution of the density distribution for $L=46a$. Specifically, two times are displayed,
    corresponding to the depositions of 0.20 ML and 2.0 ML. Thus, no stable clusters are present and the results can be
directly compared with the analytical result. Included is the theoretical stationary distribution
    (first three dominant terms from Eq. \ref{eq:stat_sol}). (b) Similar plot for $L=80a$ with $\avN = 1$.
    The distributions for the two early time steps belong to the left scale, $\rho(x\lowt{dia})$ to the
    right scale (r). The data for 0.20 ML is shown for the cluster-free ensemble (CF).
    (c) Same for $L=146a$ as in (b), using the three coverages 0.01 ML, 0.03 ML
    (averaged), and 2.0 ML.}
\end{figure}

For an analytical treatment one can calculate the time dependent particle density distribution from the
partial differential equation (PDE)
\begin{equation}
  \label{eq:pde}
 \frac{d}{dt} \rho(x,y,t) = D \Delta \rho(x,y,t) + F
\end{equation}
 with absorbing
boundary conditions \cite{Kalischewski2008}
\begin{equation}
  \label{eq:boundary_cond}
  \rho(x=0,L,y,t) = \rho(x,y=0,L,t)=0.
\end{equation}
This ansatz is based on the Burton-Cabrera-Frank theory \cite{Burton1951299}. Later
we compare the simulations with the stationary long-time solution of this PDE which can be written as
\begin{equation}
\label{eq:stat_sol}
\rho\lowt{stat}(x,y) = \sum_{m,n,\mathrm{odd}} A_{m,n}\frac{F L^2}{D} \sin \left (\frac{m\pi x}{L} \right ) \sin \left (\frac{n\pi y}{L} \right)
\end{equation}
with the numerical values $A_{m,n} = 16/(\pi^4 m n (m^2 + n^2))$ and $D$ the diffusion constant for the
random walk. Thus, $\rho\lowt{stat}(x,y)$ is dominated by the single term $m=n=1$. In what follows the maximum of $\rho\lowt{stat}(x,y)$
i.e. \mbox{$\rho\lowt{stat}(x=L/2,y=L/2)$} is correspondingly denoted as $\rho^*\lowt{stat}$.  We checked that for
$L=46$, for which no cluster-formation is observed, the numerically and analytically determined stationary density
agree very well, compare Fig. \ref{fig:fig3} and \ref{fig:fig3b} a) \cite{Vvedensky199042, Kalischewski2008}.
In particular (see Fig. \ref{fig:fig3}),
one finds $\rho^*\lowt{stat} \approx \rho^*$. The minor deviations may be related to the fact that the effective value
of the system $L$ is reduced due to the finite size of the adsorbed layer of molecules at the gold stripes. However, for bigger
$L$ this effect decreases.

To better understand the mechanisms of cluster formation we start with the discussion of $L=80a$.
In Fig. \ref{fig:fig3b} b) we show the time-dependent along diagonals averaged density $\rho(x_i,x_i)$
(denoted $\rho(x\lowt{dia})$).
For short times (coverage of 0.01 ML where no
relevant ($ < 1\%$) cluster formation has occurred,
the stationary state is not yet reached. In the
opposite limit of long times (2 ML, formation of clusters in 94.6\% of all realisations) the
density $\rho(x\lowt{dia})$ strongly increases in the centre of the system. This reflects the
presence of clusters in that region.
In order to learn about the mechanism of cluster formation we also study the case of intermediate
times (coverage of 0.20 ML) where in $35\%$ of all realisations clusters have grown. In
Fig. \ref{fig:fig3b} b) we show $\rho(x\lowt{dia})$ in the cluster-free ensemble. Studying this ensemble has
the advantage that one is at the same time sensitive to the past (no cluster growth in that subensemble)
and the future (conditions for possible future cluster growth). The
numerical data agree very well with the analytical solution $\rho\lowt{stat}(x,x)$ as shown in Fig. \ref{fig:fig3}
and \figref{\ref{fig:fig3b}} b). The nucleation process preferentially takes place in the center of the cell, where
the particle density is the highest. It is denoted as $\rho^*\lowt{form} \approx 0.016$.

Based on this observation we formulate the hypothesis that cluster growth is basically occurring for
\mbox{$\rho(x,y) \approx \rho^*\lowt{form}$}. To check this hypothesis we analyze
the case $L=146a$ for which $\rho^*\lowt{stat} > \rho^*\lowt{form}$; see Fig. \ref{fig:fig3}.
Before the actual density distribution $\rho(x,y)$ reaches the stationary distribution, the density has to
cross $\rho^*\lowt{form}$.  Exactly in this time regime nucleation sets in. E.g., for $\rho^* =  \rho^*\lowt{form}$
the nucleation probability is $40.5\%$ and approaches $88.0\%$ for $\rho^* \approx 1.4\rho^*\lowt{form} $
(see Fig. \ref{fig:fig3}). This reflects the strong dependence of nucleation
rate on density \cite{islands_mounds_atoms}. Furthermore, in this time regime there is a large area for
which $\rho(x_i,y_j) \ge \rho^*\lowt{form}$. As a consequence many nuclei can growth precisely in this area
(see Fig. \ref{fig:fig3b} c). Thus, only for
\begin{equation}
\label{eq:control}
\rho^*\lowt{stat} ( \propto L^2 F ) \approx \rho^*\lowt{form}
\end{equation}
$\rho(x,y)$ neither reaches values around $\rho^*\lowt{form}$ for a large area nor is it everywhere smaller than $\rho^*\lowt{form}$.
Thus, Eq. \ref{eq:control} is the condition for single-cluster growth, going along with good position control.
Interestingly, the long-time density
$\rho(x,y)$ for $L=146a$, reflecting the nature of the formed clusters, does not display a maximum in the middle of the cell
but rather two maxima close to the boundaries of the spatial region where cluster-formation can occur. This effect may
be explained by the fact that clusters close to these boundaries can attach the large number of freely diffusing particles
between that boundary and the gold stripe. This also rationalizes the increased cluster
size close to these boundaries, as reported also for the experimental systems.
The increased cluster size also leads to a large island size for the \SI{4.0}{\micro\meter} grid than for the
\SI{2.0}{\micro\meter} grid or on bare SiO$_2$.

\begin{figure}
  \centering
  \includegraphics[width=0.5\textwidth]{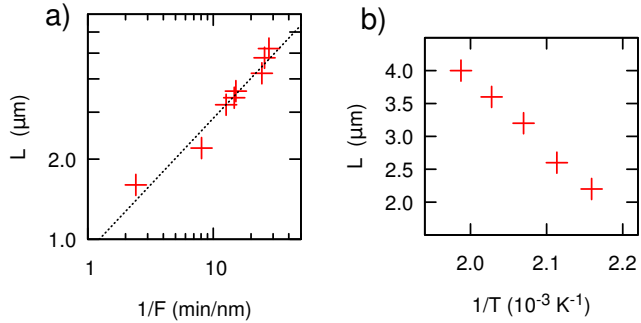}
  \caption{\label{fig:fig4}Dependence of Au grid size of one NPB island in the cell on a) double logarithmic plot of beam
    flux at substrate temperature of \SI{200}{\celsius} with straight line of the slope $1/2$ and b) substrate temperature at
    beam flux of \SI{0.12}{\nm/\minute}.}
\end{figure}

Finally, we show experimentally that by either varying the deposition flux or the temperature, one can reach
a single-cluster growth for different cell sizes; see Fig. \ref{fig:fig4}.
For a given substrate temperature of \SI{200}{\celsius}, the grid size can be changed
from \numrange{4.8}{1.6} \si{\micro\meter} varying with $F$ from
\numrange{0.04}{0.37} \si{\nano\meter/\minute}, as shown in the double logarithmic plot in Fig.
\ref{fig:fig4} a).
Based on Eq. (\ref{eq:control}) one has
$L \propto ( \rho^*\lowt{form}/F)^{1/2}$.
Assuming that $\rho^*\lowt{form}$ is independent from $F$ one gets the scaling $L \propto F^{-\frac{1}{2}} $. This seems
to be the case in our experimental
system, compare \figref{\ref{fig:fig4}} a). There we added a straight line with the slope $1/2$.
This clearly reveals a key effect of the inverse dependence of cell size and
flux to obtain single-cluster growth; see also \cite{Zuo199472}. To obtain the complete flux dependence one also has
to take into account the precise flux dependence of $\rho^*\lowt{form}$, which is, however, beyond
the scope of this work. Nonetheless the results from the step growth on vicinal surfaces
\cite{Ranguelov200775} suggest the scaling $L \propto (1/F)^{\frac{\chi}{2}}$  with
$\chi = i^*/(i^*+2)$ and $i^*$ as the critical nucleus size. This result coincides
with experimental data for a large critical nucleus size.

In Fig. \ref{fig:fig4} b) the beam flux is fixed at \SI{0.12}{\nano\meter/\minute}, the grid
size varies from \numrange{2.2}{4.0} \si{\micro\meter}, giving a linear plot
of grid size $L$ vs $1/T$.
With decreasing temperature the cluster formation becomes more
efficient so that $\rho^*\lowt{form}$ and, according to Eq. (\ref{eq:control}), $L$ decrease with increasing $1/T$.
For a detailed discussion of the temperature scaling the analyzed temperature range is too small. For general
reasons, an Arrhenius scaling is very likely to hold \cite{Ranguelov200775}.

In summary, we present a concept to control the position
of molecular aggregates by regular patterning of the
substrate with gold.
The experimentally observed and with simulations reproduced single-cluster formation is
determined by position control as well as the growth control, which leads
to the excellent short- {\it and} long-range order of the pattern. This can be understood
via Eq. \ref{eq:control} as derived from analysis of the numerical data.
Its physical background involves the strong
sensitivity of the nucleation rate on density, and the emergence of sin-type
stationary density profiles, displaying a well-defined maximum.
The single-cluster growth can be obtained for a large range of experimental
parameters. Since, furthermore, the mechanism is very general, it
is not restricted to NPB, rather the nature of the molecule in our model is only
reflected by the specific value of $D$ and $\rho^*\lowt{form}$, so it
can be directly applied to different molecules \cite{Supp_mat2}.

\begin{acknowledgements}
This work was supported through the Transregional Collaborative Research
Centre TRR 61 (projects B1 and B12) by the DFG.
\end{acknowledgements}

\end{document}